\documentclass[12pt]{article}
\usepackage{aaspp4}

\slugcomment{Submitted to {\it New Astronomy}}

\newcommand{\lil}[1]{{\rm\scriptscriptstyle #1}}
\newcommand{\mbf}[1]{\mbox{\boldmath ${#1}$}}

\newcommand{\delE}{\Delta E}
\newcommand{\delL}{\Delta L}
\newcommand{\delLo}{{\Delta L}_{\rm orb}}
\newcommand{\delvL}{\Delta \mbf{L}}
\newcommand{\Fbar}{\overline{\mbf{F}}}
\newcommand{\Lmax}{L_{\rm max}}
\newcommand{\Lmin}{L_{\rm min}}
\newcommand{\LGR}{L^\lil{GR}}
\newcommand{\rmax}{r_{\rm max}}
\newcommand{\rmin}{r_{\rm min}}
\newcommand{\rt}{r_{\rm t}}
\newcommand{\torb}{t_{\rm orb}}

\newcommand{\tprec}{t_{\rm prec}}
\newcommand{\tprecL}{t^L_{\rm prec}}
\newcommand{\tpreck}{t^{\mbf{\scriptstyle k}}_{\rm prec}}
\newcommand{\tGR}{t^{\rm \lil{GR}}_{\rm prec}}
\newcommand{\tKep}{t^{\rm \lil{Kep}}_{\rm prec}}

\newcommand{\trel}{t_{\rm rel}}

\newcommand{\dd}{\,{\rm d}}
\newcommand{\Msun}{{M_{\sun}}}
\newcommand{\ergs}{{\;{\rm erg\ s}^{-1}}}
\newcommand{\pc}{{\;{\rm pc}}}
\newcommand{\be}{\begin{equation}}
\newcommand{\ee}{\end{equation}}

\begin{document}

\title{Resonant Relaxation in Stellar Systems}

\author{Kevin P. Rauch and Scott Tremaine}
\affil{Canadian Institute for Theoretical Astrophysics, \\ University 
of Toronto, \\ 60 St. George St., Toronto M5S 3H8, Canada}
\authoremail{rauch@cita.utoronto.ca ; tremaine@cita.utoronto.ca}
\notetoeditor{FAX: (416) 978-3921}

\begin{abstract}
We demonstrate the existence of an enhanced rate of angular momentum
relaxation in nearly Keplerian star clusters, such as those found in the
centers of galactic nuclei containing massive black holes.  The enhanced
relaxation arises because the radial and azimuthal orbital frequencies in a
Keplerian potential are equal, and hence may be termed {\em resonant
relaxation\/}.  We explore the dynamics of resonant relaxation using both
numerical simulations and order-of-magnitude analytic calculations. We find
that the resonant angular momentum relaxation time is shorter than the
non-resonant relaxation time by of order $M_\star/M$, where $M_\star$ is the
mass in stars and $M$ is the mass of the central object. Resonance does not
enhance the energy relaxation rate. We examine the effect of resonant
relaxation on the rate of tidal disruption of stars by the central mass; we
find that the flux of stars into the loss cone is enhanced when the loss cone
is empty, but that the disruption rate averaged over the entire cluster is not
strongly affected.  We show that relativistic precession can disable resonant
relaxation near the main-sequence loss cone for black hole masses comparable
to those in galactic nuclei. Resonant dynamical friction leads to growth or
decay of the eccentricity of the orbit of a massive body, depending on whether
the distribution function of the stars is predominantly radial or tangential.
The accelerated relaxation implies that there are regions in nuclear star
clusters that are relaxed in angular momentum but not in energy;
unfortunately, these regions are not well-resolved in nearby galaxies by the
Hubble Space Telescope.
\end{abstract}

\keywords{black hole physics --- galaxies: active --- 
galaxies: kinematics and dynamics --- galaxies: nuclei --- stellar dynamics}

\notetoeditor{PACS: 98.10.+z ; 98.54.Cm ; 98.62.Js ; 98.62.Dm}

%
%
The dynamical evolution of galactic and active galactic nuclei (AGNs) is
determined by numerous physical processes, including two-body relaxation,
stellar collisions/mergers, tidal disruption, binary formation, stellar
evolution, and disk hydrodynamics (e.g., Spitzer \& Stone 1967; Norman \&
Scoville 1988; Quinlan \& Shapiro 1990; Murphy et al.\ 1991). In most
cases, however---and in contrast to systems such as globular clusters---the
two-body relaxation times are so long that relaxation has had little
observable effect on the nucleus so far. The slow relaxation rates in galactic
nuclei are responsible, for example, for the general failure of stellar tidal
disruption by relaxation into the loss cone (Frank \& Rees 1976; Lightman \&
Shapiro 1977) as a viable AGN fueling mechanism (Hills 1975; Frank 1978;
McMillan et al.\ 1981), and for the absence of core collapse or
mass segregation at observable resolutions in all but a few nearby galaxies
(e.g., M33; Kormendy \& McClure 1993).

Normally the gravitational potential in a galaxy is determined by the stars
and distributed dark matter. However, in galactic nuclei possessing 
massive black holes, the potential in the inner nucleus is dominated by the
black hole and hence nearly Keplerian, so that eccentric orbits maintain their
spatial orientation for many orbital periods. The aim of this paper is to
demonstrate that in Keplerian and other such ``resonant'' potentials the rate
of relaxation of angular momentum can be greatly enhanced, so that relaxation
can be important even if the energy relaxation timescale is longer than a
Hubble time.  We examine the dynamics of this ``resonant relaxation'' through
approximate analytic arguments and $N$-body simulations, and discuss its
influence on galactic nuclei.

\section{The Process of Resonant Relaxation}
\label{sec_analyt}

\subsection{Introduction}

The force field $\mbf{F}(\mbf{r},t)$ in an equilibrium $N$-body stellar
system can be divided into a mean force $\Fbar({\mbf
r})\equiv\langle \mbf{F}(\mbf{r},t)\rangle$ and a fluctuating force ${\mbf
f}(\mbf{r},t)\equiv \mbf{F}(\mbf{r},t)-\Fbar(\mbf{r})$, where
$\langle\cdot\rangle$ denotes time average. If $N\gg1$ the fluctuating force
is a Gaussian random field, and hence is completely described by the
correlation function $C_{ij}(\mbf{r}_1,\mbf{r}_2,\tau)\equiv \langle
f_i(\mbf{r}_1,t),f_j(\mbf{r}_2,t+\tau)\rangle$.  This fluctuating force
induces diffusion or relaxation of the deterministic orbits that stars would
follow if only the mean force field $\Fbar$ were
present. The relaxation time, $\trel$, is crudely defined so that the
diffusion of integrals of motion such as energy $E$ and angular momentum $L$
(per unit mass) is given by $\delE\sim E\,(t/\trel)^{1/2}$ and $\delL\sim
L\,(t/\trel)^{1/2}$.

The usual estimate of the relaxation time (Jeans 1913, 1916; Chandrasekhar
1942; Binney \& Tremaine 1987) is based on an infinite homogeneous stellar
system in which stars travel on straight-line orbits. This assumption is
plausible because each octave in spatial scale between the system size $R$ and
the much smaller scale $\rmin\sim R/N$ (the scale on which close encounters
produce $\sim 90\arcdeg$ deflections) contributes equally to the relaxation
rate, and for most of these octaves the approximations of homogeneity and
straight-line orbits are legitimate (summing the contributions from different
scales gives rise to the well-known Coulomb logarithm $\ln\Lambda\simeq
\ln(R/r_{\rm min})\simeq \ln(N)$ which appears in formulae for the relaxation
rate). One consequence of the assumption of straight-line orbits is that the
correlation function $C_{ij}(\mbf{r}_1,\mbf{r}_2,\tau)$ decays rapidly to zero
when $\tau$ exceeds $|\mbf{r}_1-\mbf{r}_2|/V$, where $V$ is a typical velocity
(for infinite homogeneous systems $C_{ij}\to 1/\tau$ as $\tau\to\infty$
[e.g. Cohen 1975]).  Our focus here is on stellar systems in which the
correlation function remains non-zero for much larger times, a condition that
occurs if most of the stars are near resonance.

If motion in the mean force field $\Fbar(\mbf{r})$ is regular, then
stellar orbits are quasi\-periodic with three characteristic frequencies
$\Omega_i$. The orbits are resonant if there are linear combinations of the
form $\sum_{i=1}^3k_i\Omega_i=0$ where the $k_i$ are small integers. The
simplest important examples are (i) spherical potentials, in which one
frequency is zero because all orbits remain in a fixed plane; (ii) Kepler
potentials, in which one additional frequency is zero because the apsis does
not precess; and (iii) the harmonic oscillator potential, in which the radial
frequency is twice the azimuthal frequency, so that the orbit shape is a
centered ellipse.

The possibility of enhanced relaxation in potentials that support many
near-resonant orbits was discussed by J. Ostriker two decades ago, in lectures
for a graduate course in stellar dynamics attended by one of us (Ostriker
1973).

\subsection{Non-resonant Relaxation}
\label{sec_nrrel}

We begin by examining relaxation in a near-Kepler potential.  Consider a
spherical volume of radius $R$ centered on a point mass $M$ and containing
$N\gg 1$ identical stars of mass $m$, where $M_\star\equiv Nm\ll M$. We assume
that the stellar orbits have random orientations and moderate eccentricities,
and that the density of stars is approximately uniform within $R$. The typical
stellar velocity is $V\sim (GM/R)^{1/2}$ and the characteristic orbital period
is $\torb\sim R/V$.  Since $M_\star\ll M$, each orbit is approximately a Kepler
ellipse, which precesses slowly on a timescale $\tprec$. If the precession is
dominated by the mean field from the other stars (rather than, say,
relativistic effects or an external tidal field), then
\be
\tprec\sim {M\over M_\star}\torb,
\label{eq:raucha}
\ee
which is much longer than $\torb$. 

The usual (non-resonant) relaxation rate can be estimated by the following
argument. Consider a volume of radius $r<R$, which typically contains $n\sim
N\,(r/R)^3$ stars. The instantaneous number of stars in this volume fluctuates
by an amount $n^{1/2}$; thus the fluctuating force on the length scale $r$ is
$f\sim Gmn^{1/2}/r^2$. The force fluctuates on a timescale $t_f\sim r/V$, and
the typical impulse that a star receives during this timescale is $\delta
v\sim ft_f\sim Gmn^{1/2}/(rV)$.  Since impulses in successive intervals of
length $t_f$ are uncorrelated, the total impulse after time $t$ is described
by a random walk,
\be
(\Delta v)^2\sim(\delta v)^2{t\over t_f}\sim{G^2m^2n\over r^3V}t\sim
{G^2m^2N\over R^3V}t\sim V^2{m^2N\over M^2}{t\over\torb}.
\ee 
Note that $\Delta v$ is independent of the scale $r$, which confirms that each
octave in scale contributes equally to the relaxation rate; thus the total
diffusion rate must be multiplied by a factor $\ln\Lambda$ which represents
the number of octaves that contribute to the relaxation.  We may define the
non-resonant relaxation time by $(\Delta v/V)^2=(t/\trel^{\rm nr})$, so that
\be
\trel^{\rm nr}\sim {M^2\over m^2 N\ln\Lambda}\,\torb.
\label{eq:trnr}
\ee

The fluctuating force changes both energy and angular momentum, at rates given
by
\begin{eqnarray}
(\delE/E)_{\rm nr}& \sim &(t/\trel^{\rm nr})^{1/2}\sim\alpha 
                      {mN^{1/2}\over M}(t/\torb)^{1/2},\nonumber
\\ 
(\delL/\Lmax)_{\rm nr}& \sim &(t/\trel^{\rm nr})^{1/2}\sim\eta_{\rm s} 
               {mN^{1/2}\over M} (t/\torb)^{1/2},
\label{eq:alpred}
\end{eqnarray}
where $E\sim V^2\sim GM/R$, $\Lmax^2\sim V^2R^2\sim GMR$, and $\alpha$ and
$\eta_{\rm s}$ are dimensionless constants that equal the square root of the 
Coulomb logarithm to within a factor of order unity.

\subsection{Resonant Relaxation in Near-Kepler Potentials} 
\label{sec_resrel}

To estimate the resonant relaxation rate, we imagine averaging the stellar
density over an intermediate timescale that is $\gg\torb$ but $\ll\tprec$.  On
this timescale each star can be represented by a fixed wire whose mass is the
stellar mass, whose shape is a Kepler ellipse, and whose linear density is
inversely proportional to the local speed in the elliptical orbit.

The gravitational potential from these wires is stationary and hence does not
lead to energy relaxation; thus
\be
(\delE/E)_{\rm res}=0.
\label{eq:erell}
\ee 
However, the wires exert mutual torques which induce angular momentum
relaxation. The typical specific torque on a wire is $T\sim N^{1/2}Gm/R$, and
fluctuates on a timescale $\sim\tprec$ as the wires precess in different
directions. Thus the characteristic change in specific angular momentum
$\delL\sim Tt$ over a timescale $t<\tprec$ is given by
\be
(\delL/\Lmax)_{\rm res}\sim \beta_{\rm s} {mN^{1/2}\over
M}(t/\torb),\qquad t\ll\tprec, 
\label{eq:trlrq}
\ee 
where $\beta_{\rm s}$ is a dimensionless constant of order unity. 

Over timescales $t\gg\tprec$, the change in angular momentum is described by a
random walk with increments $\delL/\Lmax\sim \beta_{\rm s} (mN^{1/2}/
M)(\tprec/\torb)$ over a characteristic time $\tprec$; thus,
\be
(\delL/\Lmax)_{\rm res}\sim\beta_{\rm s}{mN^{1/2}\over M}\left(\tprec\, t\over
\torb^2\right)^{1/2},\qquad t\gg\tprec;
\label{eq:genres}
\ee
if the precession is determined by the mean field of the other stars then
equation~(\ref{eq:raucha}) implies that 
\be
(\delL/\Lmax)_{\rm res}\sim 
\beta_{\rm s}\left(m\over M\right)^{1/2}(t/\torb)^{1/2},\qquad t\gg\tprec. 
\label{eq:bigrel}
\ee 
In this case the angular momentum relaxation time, defined by
$(\delL/\Lmax)_{\rm res}\allowbreak\sim(t/\trel^{\rm res})^{1/2}$, is given by
\be
\trel^{\rm res}\sim {M\over m}\,\torb,
\label{eq:trelk}
\ee
which, remarkably, is independent of the number of stars $N$. The resonant
relaxation time is shorter than the non-resonant relaxation time
(eq.~[\ref{eq:trnr}]) by a factor $(mN/M)\ln\Lambda$. If there is a range of
stellar masses, the factor $m$ in equation~(\ref{eq:trelk}) should be replaced
by $\int m^2\dd N(m)/\int m\dd N(m)$. 

There is no analog to the Coulomb logarithm in resonant relaxation, since the
relaxation is dominated by large-scale fluctuations. The absence of the
Coulomb logarithm implies that the resonant diffusion rate depends on the
overall structure of the stellar system and cannot be computed using the
assumption of local homogeneity, as is done for the non-resonant relaxation
rate.

On timescales longer than the resonant relaxation time but shorter than the
non-resonant relaxation time, a stellar system in a Kepler potential should be
in the maximum-entropy state consistent with its original total angular
momentum $\mbf{L}_{\rm tot}$ and energy distribution $N(E)$ (the number of
stars with energy $<E$, which is invariant on this timescale since there is no
resonant energy relaxation). This state is described by the phase-space
distribution function
\be
f(\mbf{r},\mbf{v})=w(E)\exp(-\mbf{b}\cdot\mbf{L}),
\label{eq:equil}
\ee
where $E={1\over2}v^2-GM/r$, $\mbf{L}=\mbf{r}\times\mbf{v}$, and $\mbf{b}$ 
and $w(E)$ are determined implicitly by the constraints 
\begin{eqnarray}
\mbf{L}_{\rm tot}&=&\int f(\mbf{r},\mbf{v})\mbf{L}\,
\dd\mbf{r}\dd\mbf{v},\nonumber \\ \noalign{\vskip -0.5em}
& & \\ \noalign{\vskip -0.5em}
{\dd N(E)\over \dd E}&=&\int f(\mbf{r}_1,{\mbf
v}_1)\delta(E-E_1)\dd \mbf{r}_1\dd \mbf{v}_1.\nonumber 
\end{eqnarray}
If $\mbf{L}_{\rm tot}=0$ then $\mbf{b}=0$ and the distribution function is a
function of energy alone (i.e., the cluster is isotropic).
If there is a range of masses then (\ref{eq:equil})
is replaced by
\be
f(\mbf{r},\mbf{v},m)=w(E,m)\exp(-m\,\mbf{b}\cdot\mbf{L}).
\ee

\subsection{Resonant Relaxation in Near-spherical Potentials} 
\label{sec_spherpot}

A more limited form of resonant relaxation is present in {\em any} spherical 
potential. Consider a generic spherical potential and average
the stellar density over a timescale $\gg\torb$. On this timescale each star
is smeared into an axisymmetric annulus whose inner and outer radii are the
pericenter and apo\-center distances.  The gravitational potential from these
annuli is stationary and hence does not lead to energy relaxation; thus, as in
equation~(\ref{eq:erell}), 
\be
(\delE/E)_{\rm res}=0.
\ee 
The annuli exert mutual torques which induce angular momentum relaxation;
however, in contrast to the Kepler case the torques are perpendicular to the
orbit normals (because the averaged orbits are axisymmetric annuli rather than
eccentric wires). Thus the vector torque $\mbf{T}_{ij}$ between two annuli
with vector angular momenta $\mbf{L}_i$ and $\mbf{L}_j$ satisfies ${\mbf
T}_{ij}\cdot\mbf{L}_i=\mbf{T}_{ij}\cdot\mbf{L}_j=0$; in other words the
torques change the directions of the angular momentum vectors but not their
magnitudes. Thus, in contrast to equation~(\ref{eq:trlrq}), there is no
resonant relaxation of the scalar angular momentum,
\be
(\delL/\Lmax)_{\rm res}=(\Delta|\mbf{L}|/\Lmax)_{\rm res}=0,
\ee
but there {\em is} resonant relaxation of the {\em vector} angular
momentum. The resonant relaxation rate may be estimated by analogy with
equation~(\ref{eq:trlrq}). The typical specific torque on an annulus is $T\sim
N^{1/2}Gm/R$, and fluctuates on a timescale $\sim\tprecL$. Here $\tprecL$ is
the precession time for the angular momentum vector ({\em not} the apsis, as
in \S\S~\ref{sec_nrrel}\ and \ref{sec_resrel}), and is determined by the 
stochastic component of the potential, $\tprecL\sim N^{1/2}\torb/\mu$.

The characteristic change in vector angular momentum over a timescale
$t<\tprecL$ is given by $|\Delta\mbf{L}|\sim Tt$.  Since we are considering
general near-spherical potentials, we no longer require that the potential is
dominated by a black hole, $M_\star\ll M$; thus the maximum angular momentum
is given by $\Lmax^2\sim V^2R^2\sim G(M+M_\star)R$ and we have 
\be
(|\Delta\mbf{L}|/\Lmax)_{\rm res} \sim \mu{\beta_{\rm v}\over N^{1/2}}
(t/\torb),\qquad t\ll\tprecL,
\label{eq:sphtr}
\ee 
where $\beta_{\rm v}$ is a dimensionless constant of order unity, and
\be
\mu={M_\star\over M_\star+M}
\label{eq:mudef}
\ee
($\sim 1$ if the potential is dominated by the stars themselves).

Over timescales $t\gg\tprecL$, the change in angular momentum is described by
a random walk. The increments $\Delta\mbf{L}$ of the random walk are
determined by evaluating (\ref{eq:sphtr}) at $t\sim\tprecL$, which yields
$|\Delta\mbf{L}|/\Lmax\sim 1$. In other words the angular momentum vectors
drift over the whole velocity sphere on a timescale of order $\tprecL$, 
implying $\trel^{\rm res}\sim \tprecL\sim N^{1/2}\torb/\mu$, which is shorter
than the non-resonant relaxation time $\trel^{\rm nr}$ (eq.~[\ref{eq:trnr}])
by a factor $(\mu/N^{1/2})\ln\Lambda$.

A closely related form of resonant relaxation is present in general
axisymmetric potentials that are nearly spherical. Once again there is no
resonant relaxation of the scalar angular momentum, but there is resonant
relaxation of the vector angular momentum. In this case only the $z$-component
of the vector angular momentum is conserved by motion in the mean potential,
so we focus on this component. By analogy with equation (\ref{eq:sphtr}) we
have
\be
(|\Delta L_z|/\Lmax)_{\rm res} \sim \mu{\beta_{\rm z}\over N^{1/2}}
(t/\torb),\qquad t\ll\tprecL,
\label{eq:sphtrz}
\ee 
where in this case $\tprecL$, the precession time for the angular momentum 
vector, is determined by the non-spherical component of the mean potential. 
Over timescales $\gg \tprecL$, the change in $L_z$ is described by 
a random walk with increments $|\Delta L_z|/\Lmax\sim 
(\mu\beta_{\rm z}/N^{1/2})(\tprecL/\torb)$ over a
characteristic time $\tprecL$; thus, 
\be (|\Delta L_z|/\Lmax)_{\rm res} \sim \mu{\beta_{\rm z}\over N^{1/2}} 
\left(\tprecL\,t\over\torb^2\right)^{1/2},\qquad t\gg\tprecL.
\label{eq:vecrelz}
\ee 
The relaxation time, defined by $(|\Delta L_z|/\Lmax)_{\rm res}\sim 
(t/\trel^{\rm res})^{1/2}$, is then given by
\be
\trel^{\rm res}\sim {N\over \mu^2}{\torb^2\over\tprecL},
\ee 
which is shorter than $\trel^{\rm nr}$
by a factor $(\torb/\tprecL)\ln\Lambda$. 

The vector angular momentum also diffuses from non-resonant relaxation, at a
rate (cf. eq.~[\ref{eq:alpred}])
\be
(|\Delta\mbf{L}|/\Lmax)_{\rm nr} \sim \mu{\eta_{\rm v}\over N^{1/2}} 
        (t/\torb)^{1/2},
\label{eq:nrvec}
\ee
where $\eta_{\rm v}$ is a dimensionless constant that equals the square root 
of the Cou\-lomb logarithm to within a factor of order unity. 

\subsection{Resonant Relaxation in Regular Potentials} 

Motion in any time-independent, regular potential can be described by
action-angle variables $(J_i,\phi_i)$ and a Hamiltonian $\overline H({\mbf
J)}$. For spherical potentials the actions $(J_1,J_2,J_3)$ can be chosen to be,
respectively, the radial action, the total angular momentum, and the
$z$-component of the angular momentum (e.g., Tremaine \& Weinberg 1984). The
motion is quasiperiodic with fundamental frequencies $\dot\phi_i=\Omega_i({\mbf
J})=\partial \overline H/\partial J_i$. In a spherical potential, $\Omega_1$
is the radial frequency, $\Omega_2$ is the azimuthal frequency, and
$\Omega_3=0$; in a Kepler potential $\Omega_1=\Omega_2$. Resonant relaxation
can be important if the resonance condition $\sum_{i=1}^3 k_i\Omega_i= 0$ is
approximately satisfied for most stars, where the $k_i$'s are small integers
with no common factor (e.g., $k_1=k_2=0$ for a spherical potential; $k_1=-k_2$
for a Kepler potential).

To isolate the effects of resonant relaxation we perform a canonical
transformation to new action-angle variables $(K_i,\psi_i)$ defined by the
generating function
\be 
S(K_i,\phi_i)=K_1\sum_{i=1}^3 k_i\phi_i +K_2\phi_2+K_3\phi_3.
\ee
Then
\be
J_1=k_1K_1,\quad J_2=k_2K_1+K_2,\quad J_3=k_3K_1+K_3,
\ee 
and
\be
\psi_1=\sum_{i=1}^3 k_i\phi_i,\quad \psi_2=\phi_2,\quad \psi_3=\phi_3.
\ee 
The resonance condition implies that $\psi_1$ is slowly varying; the
effects of resonant relaxation are therefore described by a fluctuating
Hamiltonian of the form $h(\mbf{K},\psi_1)$. The resonant Hamiltonian does
not depend on the non-resonant angles $\psi_2$ and $\psi_3$ or on the time, so
Hamilton's equations imply that the conjugate momenta $K_2$ and $K_3$ and the
total energy $E=\overline H+h$ are unaffected by resonant relaxation; this in
turn implies that the changes in the energy and actions caused by resonant
relaxation satisfy the constraints
\be
\Delta E=0,\qquad \Delta \mbf{J}=C(t)\mbf{k},
\ee 
where $C(t)$ is scalar.

Resonant relaxation leads to diffusion of $C(t)$ which is described by analogs
of equations (\ref{eq:sphtr}) and (\ref{eq:vecrelz}),
\begin{eqnarray}
(\Delta C/\Lmax)_{\rm res} &\sim &\mu {\gamma\over N^{1/2}}
(t/\torb), \qquad t\ll\tpreck;\nonumber \\ \noalign{\vskip -0.5em}
& & \\ \noalign{\vskip -0.5em}
(\Delta C/\Lmax)_{\rm res} &\sim &\mu {\gamma\over N^{1/2}}
\left(\tpreck\, t\over\torb^2\right)^{1/2}, \qquad t\gg\tpreck,\nonumber
\end{eqnarray}
where $\gamma$ is a dimensionless constant of order unity, $\mu$ is defined in
equation~(\ref{eq:mudef}), $\tpreck\sim (\sum k_i\Omega_i)^{-1}$
is the time required for the slow angle
$\psi_1$ to change by one radian, and
we have assumed that $|{\mbf k}|$ is small.

\subsection{Resonant Friction}

Chandrasekhar (1943) showed that the stochastic changes in velocity associated
with relaxation are accompanied by a systematic drag which he named
``dynamical friction.'' For massive objects (MOs) orbiting in a
stellar system, orbital evolution from dynamical friction is much faster than
evolution from stochastic relaxation. Near-resonances can enhance dynamical
friction just as they enhance stochastic relaxation, leading to an effect we
may call ``resonant friction.''

To analyze resonant friction we use the formalism developed by Lynden-Bell \&
Kalnajs (1972); the $z$-component of the specific torque on the MO from
dynamical friction is (Tremaine \& Weinberg 1984, eq. 65) 
\be T_z=4\pi^4m_0\sum_{\mbf{\scriptstyle k},k_3\ge0}k_3
\int \dd \mbf{J} \sum_i k_i {\partial
f\over\partial J_i}\left|\Psi_{\mbf{\scriptstyle k}}\right|^2 \delta({\mbf
k}\cdot{\mbf{\Omega}}-\omega_{\mbf{\scriptstyle k}}).
\label{eq:wein}
\ee 
Here $m_0\gg m$ is the mass of the MO, $f(\mbf{J})$ is the phase-space density
of bound stars, and the potential of the MO has been written in the form
\be 
U(\mbf{r},t)=m_0\,\hbox{Re}\left\{\sum_{\mbf{\scriptstyle k},k_3\ge0}
\Psi_{\mbf{\scriptstyle k}}({\mbf
J})\exp[i(\mbf{k}\cdot{\mbf{\phi}}-\omega_{\mbf{\scriptstyle k}}t)]\right\}.  
\ee 
Equation (\ref{eq:wein}) contains both resonant and non-resonant contributions
to dynamical friction.

Now let us assume that the stellar system is spherically symmetric and that
the distribution function depends only on energy and angular momentum,
$f=f(E,L)$, consistent with Jeans' theorem. Without loss of generality we may
assume that $z=0$ is the orbital plane of the MO, so that the total specific
torque on the MO is $T=T_z$. Using $\partial E/\partial J_i=\partial\overline
H/\partial J_i=\Omega_i$ and $L=J_2$ we have
\be 
T=4\pi^4m_0\sum_{\mbf{\scriptstyle k},k_3\ge0}k_3\int \dd \mbf{J}\left(
\omega_{\mbf{\scriptstyle k}}{\partial
f\over\partial E}+k_2{\partial f\over\partial L}\right)\left|
\Psi_{\mbf{\scriptstyle k}}\right|^2\delta(\mbf{k}\cdot{\mbf{\Omega}}-
\omega_{\mbf{\scriptstyle k}}).
\label{eq:weine}
\ee 
Now suppose that there is a near-resonance for the triplet ${\mbf
k}_r$. Formally, we set $\mbf{k}_r\cdot{\mbf{\Omega}}=\epsilon\,{\mbf
k}_r\cdot\widetilde{\mbf{\Omega}}$ and $\omega_{\mbf{\scriptstyle
k}_r}=\epsilon\,\widetilde\omega_{\mbf{\scriptstyle k}_r}$ and let
$\epsilon\to 0$. Keeping
only the largest terms in (\ref{eq:weine}) gives the contribution
of the triplet $\mbf{k}_r$ to the specific torque from resonant friction, 
\be 
T={4\pi^4m_0\over \epsilon}k_{3,r}k_{2,r}\int \dd \mbf{J} {\partial
f\over\partial L}\left|\Psi_{\mbf{\scriptstyle k}_r}\right|^2\delta({\mbf
k}_r\cdot\widetilde{\mbf{\Omega}}-\widetilde\omega_{\mbf{\scriptstyle k}_r}),
\label{eq:weinee}
\ee
which diverges as $\epsilon\to0$.

Evaluating expressions such as (\ref{eq:weinee}) is arduous, mostly because of
the complexity of the action-angle expansion $\Psi_{\mbf{\scriptstyle k}}$
of the potential
of a point mass (Weinberg 1986; see also Hernquist \& Weinberg 1989). However,
it is simple to derive some qualitative effects of resonant friction: 

\begin{itemize}

\item Equation (\ref{eq:weinee}) shows that there is no resonant friction 
when the distribution function is isotropic ($\partial f/\partial L=0$). The
term $\propto \partial f/\partial E$ in equation~(\ref{eq:weine}) contributes
only non-resonant friction. 

\item The resonant torque is zero if the stars are on circular orbits (in 
this case $\Psi_{\mbf{\scriptstyle k}}=0$ unless $k_1=0$, and if $k_1=0$ then
resonance requires $k_2=0$ since $\Omega_2\not=0$ and $\Omega_3=0$); similarly,
the resonant torque is zero if the MO is on a circular orbit.

\item The orbits that contribute most to the friction are those with near-zero
inclinations, since these remain closest to the zero-inclination orbit of the
MO and precess in the same direction. Zero-inclination orbits have $\Psi_{\mbf{
\scriptstyle k}}=0$
unless $k_2=k_3$ (Weinberg \& Tremaine 1984). Together with equation
(\ref{eq:weinee}), this suggests that the dominant contribution to the
resonant torque comes from terms with $k_{3,r}k_{2,r}>0$, which in turn
implies that the sign of the torque on the MO is the sign of $\partial
f/\partial L$.  In the common case where orbits are predominantly radial,
$\partial f/\partial L <0$, resonant friction removes angular momentum from
the MO orbit at constant energy, thereby increasing its eccentricity. This
effect can dominate the eccentricity evolution of black-hole binaries in
galactic nuclei and may promote the merger of binary black holes, since
emission of gravitational radiation is more efficient for an eccentric binary
(Begelman et al. 1980, Quinlan 1996).

\item The rate of growth or decay of angular momentum of the MO through 
resonant friction is approximately given by
\be 
{1\over\Lmax}\left({\dd L\over \dd t}\right)_{\rm res}\sim {T\over \Lmax}
\sim\pm {m_0 M_\star\over M^2}{\tpreck\over\torb^2}, 
\ee 
where we have assumed that the orbit of the MO has moderate eccentricity and
that $|\partial f/\partial L|\sim f/\Lmax$. In a near-Kepler potential, where
${\mbf k}_r=(1,-1,0)$ and $\tpreck$ is given by equation (\ref{eq:raucha}), we
have
\be 
{1\over\Lmax}\left(\dd L\over \dd t\right)_{\rm
res}\sim\pm{m_0\over M}{1\over\torb}.
\label{eq:fricc}
\ee 
In this case the frictional torque is---remarkably---independent of the number
of stars, although equation~(\ref{eq:fricc}) holds only when $m_0\lesssim
M_\star$. When $m_0\gtrsim M_\star$ the precession time is
$\tprec\sim(M/m_0)\torb$, so that
\be
{1\over\Lmax}\left(\dd L\over \dd t\right)_{\rm res}\sim
\pm{M_\star\over M}{1\over\torb}.
\label{eq:friccc}
\ee
The two expressions (\ref{eq:fricc}) and (\ref{eq:friccc}) can be combined,
\be
{1\over\Lmax}\left(\dd L\over \dd t\right)_{\rm res}\sim
\pm{\hbox{min}\,(M_\star, m_0)\over M}{1\over\torb}.
\ee
For comparison, non-resonant dynamical friction removes energy and angular
momentum at the slower rate 
\be 
{1\over E}\left(\dd E\over \dd t\right)_{\rm nr}\sim {1\over\Lmax}\left(\dd L\over
\dd t\right)_{\rm nr}\sim -{m_0M_\star\over M^2}{1\over \torb}.  
\ee

\end{itemize}

Another manifestation of resonant friction occurs when the potential is nearly
spherical but the mean rotation of the stars is non-zero (for example a
stellar disk) and the orbit of the MO is inclined. In this case the
distribution function $f=f(E,L,L_z)$ depends not just on $E$ and $L$ but also
on the $z$-component of angular momentum $L_z=J_3$. Because the potential is
nearly spherical, $\Omega_3\simeq0$ (i.e. inclined orbits precess slowly) so
terms with $k_1=k_2=0$ are near-resonant. The $z$-component of the torque on
an MO from the triplets $\mbf{k}=(0,0,k_3)$ is
\be
T_z=4\pi^4m_0\sum_{k_3\ge0}k_3^2\int \dd \mbf{J} {\partial f\over\partial
L_z}\left|\Psi_{0,0,k_3}\right|^2\delta(k_3\Omega_3 -\omega_{0,0,k_3}).
\ee 
If the stellar system rotates in a prograde direction, then $\partial
f/\partial L_z$ is generally positive so $T_z$ is positive; thus the resonant
friction erodes the inclination of the MO until it settles into the equatorial
plane of the stellar system. The resonant torque is zero if the stars have
precisely zero inclination (for zero-inclination stars $\Psi_{0,0,k_3}=0$
unless $k_3=0$); the rate of change of the orbital inclination $I_0$ of the MO
is given approximately by
\be 
{1\over I_0}\left(\dd I_0\over \dd t\right)_{\rm res}\sim \pm
{M_\star m_0\over M^2}{\tprec\over\torb^2}\langle I^2\rangle, 
\ee 
where $\langle I^2\rangle$ is the mean-square inclination in the disk and we
have assumed $|\partial f/\partial L_z|\sim f/\Lmax$. 

\section{Numerical Simulations}

\subsection{Methods}

Our numerical investigation of resonant relaxation utilized two complementary
approaches: an $N$-body code that integrates the orbits of individual stars
with timestep $\ll\torb$, and an ``$N$-wire'' code that follows the evolution
of Kepler ellipses with timestep $\ll\tprec$.  The $N$-body code allows us to
follow the growth of $\delE(t)$, $\delL(t)$, and $\delvL(t)$ for each star;
the $N$-wire code involves a degree of abstraction from the original problem
and does not yield $\delE(t)$, but potentially allows $\delL(t)$ and
$\delvL(t)$ to be followed for much longer times.

\paragraph{N-body simulations} These simulations are challenging
because isolating the effects of resonant relaxation requires following the
$N$-body system for very long times (our simulations ran for up to
$10^6\,\torb$).

The $N$-body system contained three components:

\begin{itemize}

\item A fixed spherical potential, $\Phi(r)$. To measure the importance of
resonant relaxation we compared the evolution of pairs of systems that were
identical except for the fixed potentials. One potential was Keplerian,
$\Phi(r)=-GM/r$, in which there is no orbital precession, and the other was an
isochrone potential, $\Phi(r)=-GM/[b+(b^2+r^2)^{1/2}]$ (H\'enon 1959), where
$b$ is a ``core'' radius, in which the precession time is comparable to
the orbital period for $r\sim b$. We used units in which $G=M=b=1$.

\item $N$ identical ``background'' stars of mass $m$, which felt only 
the gravitational force from the fixed potential, not from each other or from
the test stars. The total mass of the background stars was much smaller than
the mass associated with the fixed potential; typically
$M_\star/M\sim10^{-2}$--$10^{-5}$.

\item $n\ll N$ identical ``test'' stars of mass $m$, which felt the
gravitational force from the central potential, the background stars, and from
one another.  (We also conducted a few simulations using test stars of unequal
masses.)  The use of separate background and test stars reduces the force
calculation to $O(nN)$ rather than the much larger $O(N^2)$ for the fully
self-consistent calculation; typically, $n/N\lesssim 0.1$.

\end{itemize}

To reduce integration errors during close encounters, the gravitational force
between stars was softened. The softening length was usually $\sim 1\%$ of the
system size; the influence of softening on the results is discussed in
\S~\ref{sec_simres}.

The initial distribution function of the stars was isotropic, with
$\dd N(E)/\dd E\propto E^{-2}$ for $E_{\rm min}<E<E_{\rm max}=10E_{\rm min}$ and
zero otherwise. For the isochrone potential, $|E_{\rm min}|= 0.05\,GM/b$; this
implies a stellar density $\rho(r)\propto r^{-2}$ for $b\lesssim r\lesssim
10\,b$. For the Kepler potential, $E_{\rm min}$ was chosen so that the typical
star had the same orbital period as in the isochrone potential.  Test stars
were chosen such that their orbits remained within the body of the cluster, so
as to avoid edge effects due to the cluster's finite radial extent; in
practice this meant limiting possible test stars to orbits that were not too
eccentric, $e\lesssim 0.8$. For example, in the isochrone potential test stars
were required to have turning points $\rmin>0.8\,b$ and $\rmax<9\,b$, implying
$e<0.84$; background stars were present in the larger region $0.2\,b<r<20\,b$.

The integration algorithm was symplectic to eliminate spurious dissipation
over the long integration times. We used either a second-order scheme
(``leap\-frog'') or its fourth-order generalization\footnote{The leapfrog map
with timestep $h$ is $H_2(h):(\mbf{r},\mbf{v})\to (\mbf{r}',\mbf{v}')$,
where $\mbf{r}_1=\mbf{r}+{1\over 2}h\mbf{v}$, $\mbf{v}'=\mbf{v}-h{\mbf
\nabla} U(\mbf{r}_1)$, $\mbf{r}'=\mbf{r}_1+{1\over 2}h\mbf{v}'$; the
fourth-order map (e.g., Yoshida 1990) is the composition of three leapfrog
steps, $H_4(h)=H_2(ah)H_2(bh)H_2(ah)$, where $a=1/(2-2^{1/3})$ and
$b=1-2a=-2^{1/3}/(2-2^{1/3})$.}, depending on the situation---the higher-order
scheme was more accurate but the lower-order scheme was more robust during
close encounters.  An additional advantage of these algorithms is that they
exactly conserve angular momentum when integrating motion in a central
potential.

The advantages of symplectic schemes are generally lost if the timestep is
varied. Thus each orbit was integrated with a fixed timestep. However, because
of the wide range of orbital periods, it was useful to allow different
timesteps for different orbits. To avoid having to interpolate when computing
forces, the different timesteps must be commensurate;  in the case of
second-order leapfrog, this is easily arranged.  Suppose that a timestep less
than $h_i$ is needed to maintain the desired accuracy for star $i$, and let
$h=\min_ih_i$. In a leapfrog scheme with step $h$ the forces are evaluated at
times $(k+{1\over2})h$, $k=0,1,2\ldots$, that is, in the middle of the
timestep. Therefore we can ensure that the timesteps are commensurate by
choosing timesteps $h_i'$, where $h_i'$ is the largest odd multiple of $h$
less than $h_i$. The further restriction that $h_i'/h$ be a multiple of 3
ensures that all of the timesteps end at the same time as the largest
timestep, making synchronized output easy. Using this approach speeded up the
$N$-body code by a factor $\sim 2$--$3$. In the fourth-order scheme, however,
forces are computed at irrational (and hence unalignable) fractions of the
interval, so that in this case the same timestep $h$ was used for all stars;
for fractional energy accuracies $\epsilon\lesssim 3\times 10^{-5}$, the
savings of the fourth-order scheme outweighed the cost of a single timestep
and produced the faster code.

The size of the timestep (in units of the radial orbital period) required to
maintain relative energy accuracy $\epsilon$ was $h\approx
(\rmin/\rmax)^{1.25} \epsilon^{1/2}$ for the second-order scheme and $h\approx
0.5(\rmin/\rmax)^{1.25} \epsilon^{1/4}$ for the fourth-order scheme, where
$\rmin$ and $\rmax$ are the orbit's turning points ($\epsilon$ is the
amplitude of the spurious energy oscillation caused by the integration
algorithm; there is no systematic energy drift, because the algorithm is
symplectic.). A useful check on the accuracy of the code is to follow a
cluster whose background stars are artificially held fixed at their initial
positions; since the test stars are then orbiting in a fixed (albeit
non-spherical) potential, their orbital energy should be conserved. This test
was mainly used to ensure that the error accumulated during close encounters
was negligible. A second constraint on the required accuracy is that the
natural orbital precession should not be artificially accelerated by
integration errors. The simulations generally used $\epsilon=10^{-6}$.

\paragraph{N-wire simulations} The heart of these simulations is an
efficient algorithm to compute the time-averaged torque between two Kepler
orbits, which was supplied to us by Jihad Touma (Touma \&
Tremaine 1996). The evolving angular momentum for each test star was computed
by straightforward integration of the time-dependent torque on the orbits
using a standard (non-symplectic) adaptive ODE integrator. 

The initial conditions and the division into test and background stars were
the same as in the $N$-body code. Of course, since the analysis assumes that
the orbits are approximately Keplerian, comparative integrations in the
isochrone potential were not possible. The simulations also do not yield
estimates for $\delE$, which is zero in this approximation. The potential
advantages of the $N$-wire simulations are that (i) they provide an
independent check on the $N$-body code; (ii) they isolate the effects of
resonant relaxation without additional ``noise'' from non-resonant relaxation;
(iii) the evolution of $\delL$ and $\delvL$ can be traced for longer times
than in the $N$-body code, due to the larger timesteps that can be used (but
see below).

\subsection{Simulation Results}
\label{sec_simres}

\begin{figure}
\plotone{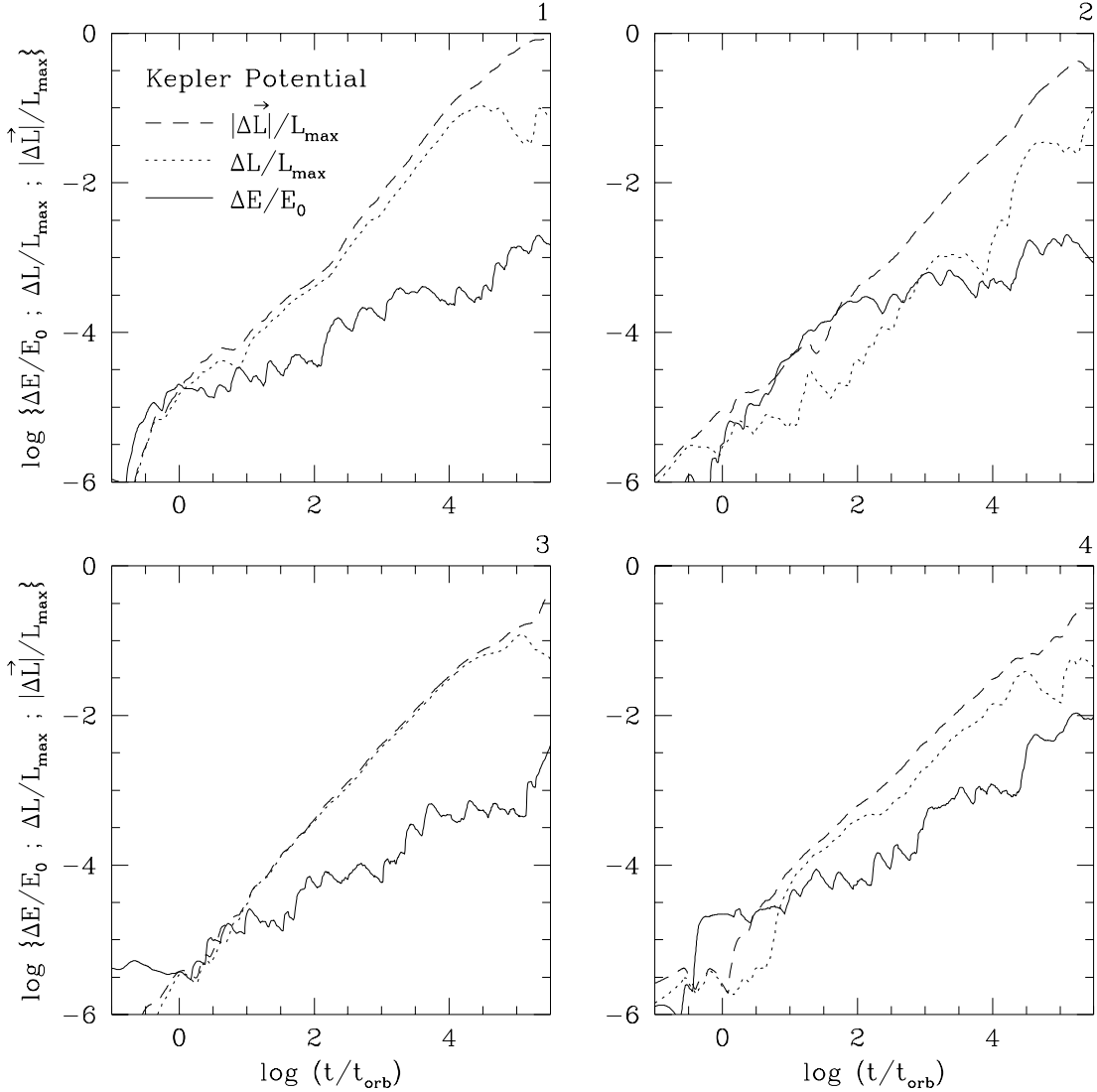}

\caption{Energy and angular momentum relaxation over time (in units of the
mean orbital time, $\torb$) for four test stars in a Keplerian star cluster.
Resonant relaxation causes $\delL$ (dotted line) and $|\delvL|$ (dashed line)
to grow almost linearly with time, in contrast to the square-root growth of
$\delE$ (solid line), which is consistent with a random walk. Note the (weak)
evidence for a turnover in the $\delL$ curves at $t\sim \tprec\approx 10^{4.7}
\torb$, where resonant relaxation is expected to become a random walk (see
eqs. \ref{eq:trlrq} and \ref{eq:bigrel}).\label{kepfig}}
\end{figure}

\begin{figure}
\plotone{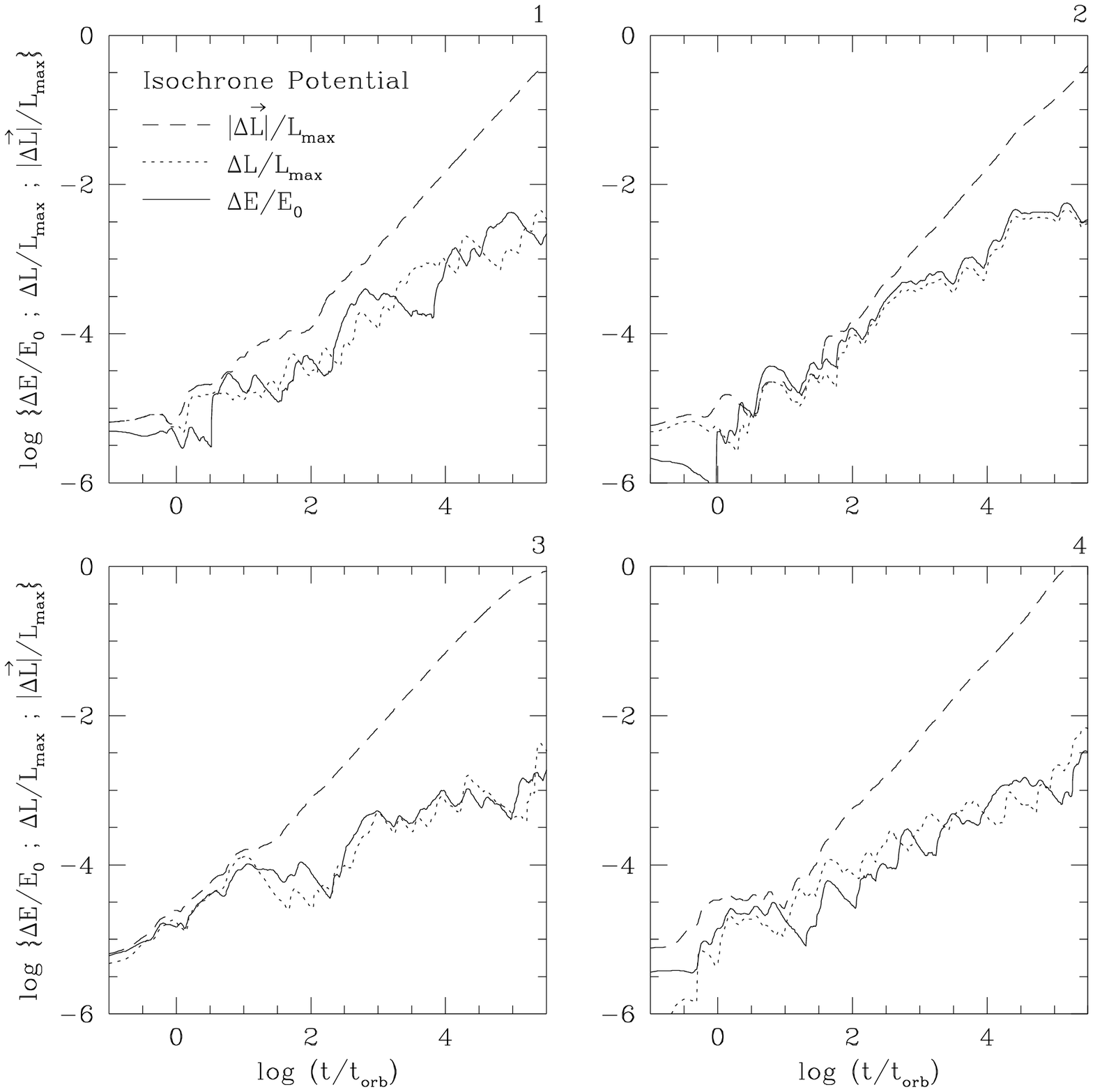}

\caption{As Figure~\ref{kepfig}
but for an isochrone potential, in which there is rapid
precession. Note that in this case $\delL/L\sim\delE/E\propto
t^{1/2}$ at all times; because the isochrone potential is spherical, however,
resonant relaxation is present in $|\delvL|$ (\S~\ref{sec_spherpot}).
\label{isofig}}
\end{figure}

A sample pair of simulations illustrating the presence of resonant relaxation
is shown in Figures~\ref{kepfig} and \ref{isofig},
which show simulations in the Kepler and
isochrone potentials respectively; we expect resonant relaxation to be
important in the former case and not the latter.  Both simulations used $n=4$
test stars out of $n+N=64$ total stars, each of mass $m=3\times 10^{-7}M$.
(Although the total number of stars in each cluster was small, the results
they display are consistent with those from larger simulations; in visual
terms the small-$N$ runs show the effects of resonant relaxation most clearly
since they can be evolved longer.)

The four panels in each of Figures~\ref{kepfig} and \ref{isofig}
plot the three quantities
$\delE/E_0$, $\delL/\Lmax$, and $|\delvL|/\Lmax$ for the four test stars,
where $E_0$ is the initial energy and 
\be 
\Lmax(E)=\cases{GM/(2|E|)^{1/2},&
Kepler potential;\cr GM/(2|E|)^{1/2}(1-2b|E|/GM),& isochrone potential,\cr}
\ee 
is the angular momentum of a circular orbit with energy $E$. The times are
given in units of $\torb$, defined as the mean radial orbital period of all
stars in the cluster.  The difference between the two potentials is clear: in
the Kepler case (Fig. \ref{kepfig}),
growth of $\delL$ systematically outpaces that of
$\delE$, by 1--2 orders of magnitude in the course of the integration, whereas
in the isochrone potential (Fig. \ref{isofig})
$\delL/\Lmax\sim \delE/E$ at all times.
In other words, if we define the (scalar) angular momentum relaxation time
$t_L$ by $\delL(t_L)\sim \Lmax(E_0)$ and the energy relaxation time $t_E$ by
$\delE(t_E)\sim E_0$, then $t_L\ll t_E$ when precession is slow, while
$t_L\sim t_E$ otherwise. In both potentials, the vector angular momentum
relaxation time $t_{\mbf{\scriptstyle L}}\ll t_E$ (cf. \S~\ref{sec_spherpot});
however, this rapid
relaxation of $\mbf{L}$ is less interesting since it reflects changes in the
orientation rather than the shape of the orbit, which are of little
consequence in non-rotating spherical clusters.

Note that for these simulations, $\tprec/\torb\sim M/M_\star\approx 10^{4.7}$,
so the figures mostly illustrate the regime $t\ll\tprec$ in which $\delL$ is
expected to increase linearly with time (eq.~\ref{eq:trlrq}). At times
$t\gg\tprec$, $\delL$ should increase as the square root of the time
(eq.~\ref{eq:bigrel}). There is a suggestion of such a turnover in Figure
\ref{kepfig}; we comment further on this issue below.

\begin{deluxetable}{lll}
\tablecaption{Parametric Relaxation Curve Models\tablenotemark{a}
  \label{paradefs}}

\tablehead{
\colhead{$\delE(\tau)/E_0$}& \colhead{$\delL(\tau)/\Lmax$}& 
\colhead{$|\delvL(\tau)|/\Lmax$}
}

\startdata
  (a)\  $\zeta m N^{1/2} \tau^\nu$&
  (c) \  $\gamma_{\rm s} m N^{1/2} \tau^{\delta_{\rm s}}$ &
  (f)\  $\gamma_{\rm v} m N^{1/2} \tau^{\delta_{\rm v}}$\nl
  (b)\  $\alpha m N^{1/2} \tau^{1/2}$&
  (d)\  $\eta_{\rm s} m N^{1/2} \tau^{1/2}+\beta_{\rm s} m N^{1/2} \tau$&
  (g)\  $\eta_{\rm v} m N^{1/2} \tau^{1/2}+\beta_{\rm v} m N^{1/2} \tau$\nl
  & (e)\  $c\tau^{b_1}/[1+(\tau/\tau_{\rm b})^{b_3}]^{(b_1-b_2)/b_3}$& \nl
\enddata

\tablenotetext{a}{ $\tau\equiv (t/\torb)$; $M=1$.}
\end{deluxetable}

We may test the analytic analysis of \S~\ref{sec_analyt}
in more detail by performing
least-squares fits on the simulation data.  We fit each of the curves in
Figures~\ref{kepfig} and \ref{isofig}---and
analogous curves for other simulations---to the variety
of power laws, sums of power laws, and broken power laws defined in Table
\ref{paradefs}. Formula (a) tests whether $\Delta E(t)$ exhibits the $t^{1/2}$ dependence
predicted by equation~(\ref{eq:alpred}), and formula (b) finds the best-fit
value of the parameter $\alpha$ in equation~(\ref{eq:alpred}). Formula (c)
tests whether $\delL(t)$ is dominated by the $t^{1/2}$ behavior characteristic
of non-resonant relaxation (eq.~\ref{eq:alpred}) or the linear growth
characteristic of resonant relaxation when $t\ll\tprec$ (eq.~\ref{eq:trlrq});
formula (d) fits simultaneously for both resonant and non-resonant relaxation
in $\delL$ (eqs. \ref{eq:alpred} and \ref{eq:trlrq}); and formula (e) fits to
a broken power law ($\propto t^{b_1}$ for small $t$ and $\propto t^{b_2}$ for
large $t$, the transition occurring at $t/\torb\sim \tau_{\rm b}$) to look for
the turnover from linear growth (eq.~\ref{eq:trlrq}) to square-root growth
(eq.~\ref{eq:bigrel}) that is expected at $t\sim\tprec$. Finally, formulae (f)
and (g) are the vector analogs of (c) and (d) (cf. eqs. \ref{eq:sphtr}\ and
\ref{eq:nrvec}).

In performing the fits, random noise in the relaxation curves of individual
test stars is a significant concern. Note first that fitting to the raw output
is undesirable because curves such as $\Delta E(t)$ cross zero many times,
which in log-log space implies passage through $-\infty$; what is wanted is a
fit to the more steadily growing {\em envelope}, ignoring the zero
crossings. This was done by smoothing the individual curves using an averaging
window of width $\Delta\log t\sim 0.1$ to replace each data point by the
weighted average of its neighbors, larger data points receiving greater weight
(the weighting factor was proportional to the square of the data point).  This
window was large enough to remove the sharp ``dropouts'' due to zero
crossings, but too small to affect the global shape of the envelope
(decreasing the window size to $\Delta\log t\sim 0.03$ changed power-law
slopes by $\lesssim 1\%$, but noticeably increased the random noise). As these
smoothed curves still contain (real!)  random variations, more robust results
were obtained by averaging the curves for all test stars in a particular
simulation, scaled temporally by their respective orbital periods, to create
composite average relaxation curves for the cluster, model parameters being
calculated from these composites.  Conceptually one can think of these curves
as belonging to a fictitious ``average'' star with orbital period $\torb$; we
will use this term to label such fits.  An alternative procedure is to derive
best-fit parameters for individual stars and then average these results; this
method is less robust since it attempts to remove the random noise {\em after}
performing the fits, which can be ineffective if this noise causes the fits to
be poor.  Therefore, all of our quantitative results are based solely on fits
to the average star as defined above.

\begin{deluxetable}{cccccccc}
\tablecaption{Best-fit Parameters for Figure~\ref{kepfig}
  (Kepler Potential)\label{keptab}}

\tablehead{
\colhead{} & \multicolumn{7}{c}{Formula Number and Parameters} \\
\tablevspace{1ex} \cline{2-8}\\ \tablevspace{-2ex}
\colhead{} & \colhead{(b)} & \colhead{(a)} & \colhead{(c)} & \colhead{(f)} &
    \colhead{(d)} & \colhead{(g)} & \colhead{(e)} \\
\colhead{Star} & \colhead{$\alpha$} & \colhead{$\zeta$, $\nu$} &
    \colhead{$\gamma_{\rm s}$, $\delta_{\rm s}$} &
    \colhead{$\gamma_{\rm v}$, $\delta_{\rm v}$} &
    \colhead{$\eta_{\rm s}$, $\beta_{\rm s}$} &
    \colhead{$\eta_{\rm v}$, $\beta_{\rm v}$} &
    \colhead{$\log\tau_{\rm b}$, $b_2$}
}

\startdata
  1& 1.73&
    3.8, 0.39&
    2.0, 0.92&
    2.4, 0.96&
    2.8, 0.92&
    3.2, 1.51&
    4.7, $-$0.17\\

  2& 5.40&
    9.9, 0.40&
    0.9, 0.84&
    3.4, 0.94&
    1.2, 0.22&
    2.5, 2.03&
    3.8, \phs1.05\\

  3& 2.31& 
    2.9, 0.46&
    2.2, 1.00&
    2.5, 1.00&
    0.0, 2.50&
    0.0, 2.76&
    4.5, $-$0.50\\

  4& 6.36& 
    5.2, 0.53&
    2.4, 0.88&
    5.6, 0.87&
    0.6, 1.17&
    4.0, 2.20&
    3.8, $-$0.10\\

  mean& 3.26&
    4.2, 0.45&
    2.0, 0.95&
    3.2, 0.95&
    1.9, 1.28&
    2.2, 2.10&
    4.4, \phs0.07\\ \tablevspace{1ex}

  AS& 3.51&
    5.2, 0.44&
    1.8, 0.90&
    3.5, 0.93&
    1.6, 0.85&
    2.4, 2.04&
    4.5, \phs0.18\\
\enddata

\end{deluxetable}

\begin{deluxetable}{cccccccc}
\tablecaption{Best-fit Parameters for Figure~\ref{isofig} (Isochrone Potential)
  \label{isotab}}

\tablehead{
\colhead{} & \multicolumn{7}{c}{Formula Number and Parameters} \\
\tablevspace{1ex} \cline{2-8} \\ \tablevspace{-2ex}
\colhead{} & \colhead{(b)} & \colhead{(a)} & \colhead{(c)} & \colhead{(f)} &
    \colhead{(d)} & \colhead{(g)} & \colhead{(e)} \\
\colhead{Star} & \colhead{$\alpha$} & \colhead{$\zeta$, $\nu$} &
    \colhead{$\gamma_{\rm s}$, $\delta_{\rm s}$} &
    \colhead{$\gamma_{\rm v}$, $\delta_{\rm v}$} &
    \colhead{$\eta_{\rm s}$, $\beta_{\rm s}$} &
    \colhead{$\eta_{\rm v}$, $\beta_{\rm v}$} &
    \colhead{$\log\tau_{\rm b}$, $b_2$}
}

\startdata
  1& 2.36&
    1.7, 0.56&
    2.3, 0.52&
    2.3, 0.85&
    2.5, 0.00&
    3.7, 0.55&
    \phs0.1, 0.52\\

  2& 5.06&
    3.0, 0.59&
    2.0, 0.61&
    1.8, 0.91&
    3.1, 0.01&
    2.1, 0.75&
    $-$0.1, 0.61\\

  3& 2.87& 
    8.5, 0.34&
    7.0, 0.34&
    4.4, 0.89&
    3.4, 0.00&
    6.5, 1.47&
    \phs0.6, 0.33\\

  4& 1.65& 
    2.7, 0.43&
    3.1, 0.44&
    1.5, 0.95&
    2.3, 0.00&
    2.6, 0.82&
    \phs0.5, 0.44\\

  mean& 2.80&
    3.5, 0.48&
    2.8, 0.50&
    3.0, 0.89&
    2.6, 0.00&
    4.6, 0.99&
    \phs0.2, 0.49\\ \tablevspace{1ex}

  AS& 2.84&
    3.2, 0.48&
    3.1, 0.48&
    2.5, 0.89&
    2.9, 0.00&
    3.6, 0.84&
    $\phantom{-}$0.5, 0.46\\
\enddata

\end{deluxetable}

Tables~\ref{keptab} and \ref{isotab}
show derived parameter values for the relaxation curves shown
in Figures~\ref{kepfig} and \ref{isofig}.
Fits to the individual test stars, the averages of the
fit parameters (labeled mean), and fits to the average star (labeled AS)
are given. Comparison of Tables~\ref{keptab} and \ref{isotab} shows the effect
of resonant relaxation quite clearly. In Table~\ref{keptab} (Kepler model)
the exponent $\delta_{\rm
s}$ for $\delL/\Lmax$ (formula [c]) is near unity, corresponding to the linear
growth expected for resonant relaxation; while the exponent $\nu$ for
$\delE/E_0$ (formula [a]) is near 0.5, corresponding to a random walk
dominated by non-resonant relaxation (recall that there is no resonant energy
relaxation). On the other hand, in Table~\ref{isotab} (isochrone model) there
is no systematic difference between the exponents $\delta_{\rm s}$ and $\nu$.

The difference between the Kepler and isochrone potentials is also reflected
in the fits to formula (d), which decomposes the relaxation into terms
$\propto t^{1/2}$ (non-resonant) and $\propto t$ (resonant). The strength of
the resonant term, which is proportional to $\beta_{\rm s}$, is comparable to
the strength of the non-resonant term in the Kepler potential, but is
essentially zero in the isochrone model.

The presence of resonant relaxation of the vector angular momentum in
spherical potentials is illustrated by the exponent $\delta_{\rm v}$ for
$\delvL/\Lmax$, which is near unity in both Tables~\ref{keptab} and
\ref{isotab} (eq.\ \ref{eq:sphtr}).

A glance at the derived values for $\log\tau_{\rm b}$ and $b_2$ in Table
\ref{keptab}
illustrates the advantage of using the average star to compute parameters
instead of averaging the fits to individual stars; the individual fits are so
noisy that averaging after the fact is nearly meaningless. Note
that $\log\tau_{\rm b}$ for the average star in Table~\ref{keptab}
is of order the
expected value, $\log(\tprec/\torb)\sim\log(M/M_\star)\sim4.7$.

\begin{deluxetable}{cccc}
\tablecaption{Global Average Best-fit Parameters\tablenotemark{a}
  \label{aveparas}}

\tablehead{
\colhead{Formula} & \colhead{Parameters} & \colhead{Kepler Potential} &
\colhead{Isochrone Potential}
}

\startdata
(b) & $\alpha$&
  3.09(0.11)\phantom{,$^{\rm b}$ 0.00(0.00)}\  [0.25]& 
  2.76(0.11)\phantom{, 0.00(0.00)}\  [0.23]\\

(a) & $\zeta,$ $\nu$&
  3.24(0.17),\phantom{$^{\rm b}$} 0.49(0.01)\  [0.23]&
  2.93(0.18), 0.49(0.01)\  [0.21]\\

(c) & $\gamma_{\rm s},$ $\delta_{\rm s}$&
  1.35(0.08),\phantom{$^{\rm b}$} 0.85(0.02)\  [0.26]&
  2.87(0.12), 0.49(0.01)\  [0.21]\\

(f) & $\gamma_{\rm v},$ $\delta_{\rm v}$&
  2.78(0.18),\phantom{$^{\rm b}$} 0.88(0.01)\  [0.17]&
  2.73(0.15), 0.84(0.02)\  [0.33]\\

(d) & $\eta_{\rm s},$ $\beta_{\rm s}$&
  1.37(0.11),\phantom{$^{\rm b}$} 0.53(0.06)\  [0.22]&
  2.76(0.09), 0.00(0.01)\  [0.24]\\

(g) & $\eta_{\rm v},$ $\beta_{\rm v}$&
  2.08(0.26),\phantom{$^{\rm b}$} 1.79(0.12)\  [0.11]&
  3.50(0.10), 0.71(0.07)\  [0.13]\\

(e) & $\log\tau_{\rm b},$ $b_2$&
  0.02(0.27)\tablenotemark{b} , 0.49(0.13)\  [0.04]&
  0.73(0.14), 0.49(0.01)\  [0.04]\\
\enddata

\tablenotetext{a}{ The numbers in parentheses are 1--$\sigma$ errors on the
best-fit parameters, which were derived by assuming equal
error (in dex) for each simulation data point and choosing this error
so that $\chi^2=1$ per degree of freedom (the required errors in the data
points in dex are given in square brackets). }

\tablenotetext{b}{ This number is $\log(\tau_{\rm b}/\tau_{\rm prec})$, not
$\log\tau_{\rm b}$.}

\end{deluxetable}

We originally hoped that the $N$-wire code could be run for many more orbital
times than the corresponding $N$-body calculation, since the required
integration step is much longer. This hope was frustrated by close encounters
of the wires, which required each wire to be divided into hundreds or even
thousands of segments for accurate calculation of the torque (the calculations
for each segment taking $\sim 10^2$--$10^3$ times longer than the simple force
calculations of the $N$-body code). These encounters dominated the computation
time, so that the $N$-wire code was significantly {\em slower} than the
$N$-body code, particularly for larger $N$. Nevertheless, the $N$-wire
simulations that were performed showed the same qualitative features as their
$N$-body counterparts, including clear evidence of resonant relaxation of
similar strength, and the suggestion of a turnover near $t\sim \tprec$. The
$N$-wire code could be sped up in several ways: by softening the potential
from the wires, by using a finer mesh close to the crossing point, or by
analytic evaluation of the torque when the wires are close to
crossing. However, as there appeared to be no new features in the relaxation
curves, a substantial effort to improve the speed was not deemed worthwhile,
and quantitative results from the wire simulations will not be given.

Table~\ref{aveparas}
lists the best-fit model parameters averaged over $\approx 30$
$N$-body simulations spanning several orders of magnitude in $m$ (from
$3\times 10^{-8}$ to $3\times 10^{-5}$) and $N$ (from 64 to 8192). The
contribution of each simulation to the average was weighted by the standard
error in the fit, which in turn was determined by choosing the error for each
data point in the simulation so that $\chi^2=1$ per degree of freedom.

The results nicely confirm our theoretical expectations: (i) In both the
Kepler and isochrone potentials, the evolution of $\delE$ is consistent with a
random walk, that is, the exponent $\nu=0.5$ to within the errors; moreover,
the rate of energy relaxation is nearly identical in the two models (the
values for $\alpha$ differ by only 10\%), which confirms that they are close
analogs except for the resonance in the Kepler case. (ii) In the isochrone
potential, the evolution of $\delL$ is consistent with a random walk, that is,
$\delta_{\rm s}=0.5$ to within the errors; while in the Kepler case $\delL$
grows much more rapidly (the derived value $\delta_{\rm s}=0.85\pm 0.02$ is
less than the expected value of 1, presumably because small-scale fluctuations
in the torque add a random-walk component---this issue is discussed further
below). There is also no evidence for a linear component in $\delL$ in the
isochrone model ($\beta_{\rm s}=0$).
(iii) The growth of $\delvL$ is more rapid than
a random walk in both potentials; once again, the derived values $\delta_{\rm
v}=0.88\pm0.01,\,0.84\pm0.02$ are lower than unity because of small-scale
fluctuations in the torque. (iv) The value of $\tau_{\rm b}$ is close to the
expected value $\tau_{\rm prec}$ in the Kepler potential ($\log(\tau_{\rm
b}/\tau_{\rm prec})=0.02\pm0.27$); moreover, as predicted by equation
(\ref{eq:bigrel}), the evolution of $\delL$ for $t\gg\tprec$ is consistent
with a random walk (which requires $b_2=0.5$; the observed value is
$0.49\pm0.13$).

Note that the diffusion in angular momentum, as measured by $\eta_{\rm s}$ and
$\eta_{\rm v}$, is stronger in the isochrone than the Kepler potential
(perhaps because the encounter velocities are smaller in the isochrone core),
while the resonant relaxation in $\mbf{L}$, as measured by $\beta_{\rm v}$, is
stronger in the Kepler potential because of the contribution to the torque
from the eccentricity of the orbits. 

We now comment on whether the linear/random-walk model (formulae [d] and [g])
provides a better fit to $\delL$ and $\delvL$ than a simple power-law
(formulae [c] and [f]). Table~\ref{aveparas}
shows that in all three cases of interest
(Kepler $\delL$ and $\delvL$, and isochrone $\delvL$) formula (d) fits better
than (c) and (g) fits better than (f), the required errors in the data points
to achieve a given $\chi^2$ being smaller. To determine whether this
difference is significant, paired sample Student's t-tests were performed to
compute the confidence level at which the hypothesis of the corresponding
$\chi^2$ values having the same mean could be excluded. For both potential
models, the resulting confidence for $\delvL$ was over $99.9\%$---the
power-law model is clearly inferior to the linear/random-walk model.  In the
case of $\delL$ (Kepler potential), applying the t-test to the full simulation
set gave a confidence of only $\approx 85\%$ that the linear/random-walk model
provides a better fit; this weaker result is not surprising since the
relaxation curves are noisier. By performing the test using only the 6
simulations with at least 32 test stars, we obtained a stronger confidence
level of $\approx 95\%$.

We also checked the scaling of the relaxation rate with $m$ and $N$, by
fitting the simulations to a formula of the form $\delE/E_0=\alpha m^a N^b
\tau^{1/2}$; this gave $\alpha=3.6\pm 0.9$, $a=1.02\pm0.01$, and
$b=0.51\pm0.03$. Thus the exponents are consistent with their predicted values
$a=1$, $b={1\over2}$ to within a few percent. Similar results were obtained
from fits to $\delL$ and $\delvL$.

We note that softening is expected to affect the strength of the random walk
terms in the simulations by changing the value of the ``$\ln \Lambda$'' factor
which these terms implicitly contain. In general, one expects $\Lambda\sim
R/\max(s, GM/v^2)$, where $s$ is the softening length and $R$ the system size.
Since $s\gg GM/v^2$ in all of our simulations, we expect that $\alpha^2\propto
t_E^{-1}\propto \ln[K\,(R/s)]$ for some factor $K\sim 1$.  To test this
scaling, we ran a series of Kepler simulations in which the softening length
was reduced by a factor of 100 (from $s\simeq 10^{-2}\, R$ to $s\simeq
10^{-4}\,R$), which should increase $\alpha$ by a factor of about $1.4$.  It
was found that $\alpha=5.5\pm 0.2$ for these runs, a factor of 1.8 larger than
in Table~\ref{aveparas}, which is in line with expectations given the
uncertainties. Similar increases were seen in other parameters, such as
$\eta_{\rm s}$ which increased by a factor of $1.6$. This observation also
supports our claim that the $\eta$ parameter really is measuring a random-walk
component in $\delL$.

\section{Implications for Galactic Nuclei}

\subsection{Black Hole Fueling Rates}
\label{sec_newtloss}

Tidally disrupted stars can be an important fuel source for massive black
holes in stellar systems, disruption occurring if the stars pass closer to the
black hole than their tidal radius, $\rt$. The pericenter of an orbit with
energy $E\gg-GM/\rt$ is within the tidal radius if its angular momentum is
less than $\Lmin\simeq(2GM\rt)^{1/2}$; the region in phase space $L<\Lmin$
is known as the ``loss cone'' because stars in this region will be disrupted
in less than one orbital period, unless their orbits are perturbed out of the
loss cone before reaching pericenter.  Angular momentum relaxation provides a
steady supply of stars to the loss cone.  This mass supply rate is interesting
in the context of active galactic nuclei (AGNs), for which tidal disruption of
stars is a possible fueling mechanism. Previous studies (e.g., Duncan \&
Shapiro 1983; David et al.\ 1987a,b; Murphy et al. 1991; Polnarev \&
Rees 1994) concluded that (non-resonant) relaxation is too slow to sustain
typical AGN luminosities, unless mass loss is dominated by other mechanisms
(such as physical collisions between stars) or additional, large-scale torques
are present (as from a nuclear bar, binary black hole, or triaxial galactic
potential).  In this section we investigate whether resonant relaxation can
enhance the supply of stars to the loss cone and hence the fueling rate of
AGNs from disrupted stars.

We shall parameterize the loss rate of stars by the
dimensionless number $\lambda(E)$, which is the fraction of stars at a given
energy that are disrupted each orbital period.
Let $\delLo$ be the root-mean-square change in scalar
angular momentum in one orbital period. For non-resonant relaxation, equation
(\ref{eq:alpred}) implies
\be
{\delLo^{\rm nr}\over\Lmax}\sim\eta_{\rm s}{mN^{1/2}\over M},
\label{eq:dellnonres}
\ee
while for resonant relaxation, equation~(\ref{eq:trlrq}) implies
\be
{\delLo^{\rm res}\over\Lmax}\sim\beta_{\rm s}{mN^{1/2}\over M};
\label{eq:dellres}
\ee 
the non-resonant contribution to $\delLo$ is stronger by of order the square
root of the Coulomb logarithm.  Lightman \& Shapiro (1977) distinguish two
cases: the ``pinhole'' limit in which $\delLo\gg\Lmin$, and the ``diffusion''
limit in which $\delLo\ll\Lmin$.  First consider the pinhole limit. Here
leakage of stars into the loss cone is too small to affect the distribution
function significantly, so that the distribution function is approximately
uniform on the energy surface in phase space, even near the loss cone. Hence
the fraction of stars at a given energy that is lost per orbital period is
simply the fractional area in the energy surface occupied by the loss cone,
\be
\qquad\lambda(E)\simeq {\Lmin^2\over\Lmax^2(E)}\qquad(\delLo\gtrsim\Lmin).
\ee
In this limit the fueling rate is independent of the strength of the
relaxation, so long as $\delLo\gtrsim\Lmin$. Therefore the loss rate in this
case is unaffected by resonant relaxation, since $\delLo$ is dominated by
non-resonant relaxation.  The pinhole limit requires 
\be
{\Lmin\over\Lmax}\lesssim \eta_{\rm s}{mN^{1/2}\over M}.
\label{eq:pinddef}
\ee 
When ``$\lesssim$'' is replaced by ``$=$,'' equation~(\ref{eq:pinddef})
defines the so-called critical radius, $r_{\rm crit}$; in general the pinhole
limit applies outside the critical radius and the diffusion limit applies at
smaller radii.

Next consider the diffusion limit. In the absence of resonant relaxation the
loss rate can be determined by solving the Fokker-Planck equation
(e.g., Lightman \& Shapiro 1977; Cohn \& Kulsrud 1978), and is found to be
\be
\qquad \lambda^{\rm nr}(E)\simeq {(\delLo^{\rm nr})^2\over
\Lmax^2(E)\ln(\Lmax/\Lmin)}\qquad (\delLo\lesssim\Lmin).
\label{eq:eqdiff}
\ee  
Now assume resonant relaxation is present. Over timescales $\ll\tprec$ the
resonant torque is approximately constant, so the angular momentum drifts at a
nearly uniform rate. Since $\tprec/\torb\sim M/M_\star\gg1$ whenever
resonant relaxation is important, we may assume that $(\tprec/\torb)\,
\delLo^{\rm res}\gg\Lmin$, at least if we are not too far inside the
critical radius. Then the evolution of the angular momentum on the scale of
the loss cone resembles a steady drift rather than diffusion, so a typical
area $\sim \Lmin\delLo^{\rm res}$ is swept into the loss cone per orbital
period, and the loss rate is
\be
\lambda^{\rm res}(E)\sim {\Lmin\delLo^{\rm res}\over\Lmax^2(E)}\qquad 
(\delLo\lesssim\Lmin).
\label{eq:lossres}
\ee

The ratio of the resonant loss rate to the non-resonant loss rate in the
diffusion limit is 
\be
\qquad{\lambda^{\rm res}\over\lambda^{\rm nr}}\sim{\Lmin\delLo^{\rm res}
\over (\delLo^{\rm nr})^2}\ln(\Lmax/\Lmin)\qquad (\delLo\lesssim\Lmin).
\ee 
Since $\delLo^{\rm res}$ and $\delLo^{\rm nr}$ differ only by a logarithmic
factor (cf. eqs. [\ref{eq:dellnonres}]\ and [\ref{eq:dellres}]), the loss rate
in the diffusion limit is much larger under resonant relaxation than under
non-resonant relaxation.  (Note that formula [\ref{eq:lossres}] neglects
the relativistic effects discussed in \S~\ref{sec_grloss},
which reduce the resonant loss rate in
galactic nuclei.)  However, this enhanced loss rate does not strongly affect
the total rate of fueling of the black hole by bound stars. The reason
is that the overall flux of stars into the cusp---which in a steady state
equals the fueling rate from bound stars---is determined by the bottleneck at
the critical radius (i.e., by the slow diffusion of orbital {\em energies},
the rate of which is unaffected by resonant relaxation); the enhanced loss
rate inside $r_{\rm crit}$ will reduce the density of stars at $r<r_{\rm
crit}$ and the radial profile of the cusp of bound stars, but it will not
strongly affect the overall disruption rate. We conclude that
resonant relaxation does not significantly enhance the viability of tidal
disruption as a possible AGN fueling mechanism.

\subsection{Relativistic Effects}
\label{sec_grloss}

We have assumed so far that stellar orbits are Keplerian whenever the
black-hole mass is much larger than the mass in stars.  At small radii,
however, relativistic effects cause orbits to precess rapidly, which quenches
the resonant relaxation.  In the present case the most important relativistic
effects are the precession of periapsis and Lense-Thirring precession of the
orbital plane.  The former effect (observed in the solar system as a component
of Mercury's perihelion precession, for example) is present around both
Schwarzschild (non-rotating) and Kerr (rotating) black holes; in both cases,
the change in the argument of periapsis per orbit in the weak-field limit is
given to leading order by
\be
\Delta\omega\simeq{6\pi (GM/c^2)\over a(1-e^2)},
\ee 
where $M$ is the hole mass, $a$ and $e$ are the orbit's semi-major axis and
eccentricity, and the precession is in the direction of motion (e.g., Darwin
1961).  Around a Kerr black hole---the expected case for the nuclei of AGNs,
since disk accretion will spin up an initially non-rotating black hole
(Bardeen 1970, Thorne 1974)---orbits also undergo Lense-Thirring precession,
which in the weak-field limit appears as a precession of the orbital angular
momentum vector around the black hole's angular momentum vector. However, this
effect merely adds another (uninteresting) secular growth term to $\delvL$,
and hence is not important for the present analysis.

Let us define a relativistic precession timescale,
$\tGR=[\pi/\Delta\omega]\,\torb$, and a non-relativistic precession timescale,
$\tKep\sim (M/M_\star)\,\torb$ (eq.~\ref{eq:raucha}), where $\torb$ is the
orbital period.  Relativistic precession can be important only when
$\tGR\lesssim \tKep$.  In fact, since the relativistic precession is prograde
($\Delta\omega^\lil{GR}>0$) while the non-relativistic precession is generally
retrograde ($\Delta\omega^\lil{Kep}<0$), relativistic effects slightly {\em
enhance\/} resonant relaxation so long as $\tGR\gtrsim\tKep$.  Even when
$\tGR< \tKep$, resonant angular momentum relaxation remains important
and is described by equation~(\ref{eq:genres}) so long as $\tGR\gg\torb$,
although the relaxation time is increased by a factor $\tKep/\tGR$.

The angular momentum of a Kepler orbit is given by $L^2=GMa(1-e^2)$, so 
the relativistic precession time may be written
\be
\tGR={1\over 6}\left(cL\over GM\right)^2\,\torb.
\ee
Thus resonant relaxation is completely quenched by relativistic effects when
$L\lesssim\LGR$, where $\LGR= GM/c$. A more precise statement is that there is
a region of angular-momentum space in which resonant relaxation is completely
quenched only if
\be
\LGR\gtrsim\max(\delLo,\Lmin);
\label{eq:einst}
\ee
if $\LGR<\Lmin$ stars are disrupted before the relativistic precession time
becomes comparable to the orbital time, and if $\LGR<\delLo$ a star can relax
across the relativistic region in less than one orbit, thereby avoiding the
relativistic precession (nearly all of which occurs at periapsis).  If the
inequality (\ref{eq:einst}) is satisfied, the angular momentum diffuses
through resonant relaxation when $L\gtrsim \LGR$ and by non-resonant
relaxation when $\Lmin+\delLo \lesssim L
\lesssim \LGR$; stars will pile up in the latter region of phase space
since non-resonant relaxation is slower. Even outside this region, 
relativistic precession can dominate the precession rate and hence
reduce the rate of resonant relaxation.

The inequality (\ref{eq:einst}) requires either that the loss cone is empty
($\delLo\lesssim\Lmin$) and that $\LGR/\Lmin\gtrsim 1$, or that
$\LGR/\delLo\gtrsim 1$ if the loss cone is full; since resonant relaxation
does not affect disruption rates if the loss cone is full, the latter case is
uninteresting. A numerical value for the former condition is easily derived;
noting that $\Lmin\simeq (2GM\rt)^{1/2}$, where $\rt\approx
2r_\star(M/m)^{1/3}$ is the tidal radius of a star of mass $m$ and radius
$r_\star$, assuming $r_\star\propto m^{2/3}$ (reasonable for both low and high
mass main-sequence stars), and normalizing to solar values, we obtain
\be
{\LGR\over\Lmin}\simeq
0.3\left(M\over10^8M_\odot\right)^{1/3}\left(M_\odot\over m\right)^{1/6}.
\label{eq:grlimit}
\ee
Resonant relaxation only dominates non-resonant relaxation when
$\tprec\gtrsim 10\,\torb$ (cf. eq.~\ref{eq:relrat})---which here implies
$L\gtrsim 8\LGR$. Therefore we may assume that relativistic
precession will substantially affect the loss rate when the loss cone is empty
and $\Lmin\lesssim 4\LGR$, which implies
\be
M\gtrsim 4\times 10^7\,\left(m\over \Msun\right)^{1/2}\,\Msun.
\label{eq:emrel}
\ee
The mass estimate (\ref{eq:emrel}) is quite uncertain, but lies squarely
inside an interesting range: the estimated masses of black holes in galactic
nuclei range from $\sim 10^6\,\Msun$ to over $10^9\,\Msun$, and the Eddington
luminosity corresponding to (\ref{eq:emrel}) is $L_{\rm Edd}\sim 5\times
10^{45}\ergs$---a typical AGN luminosity.

To summarize, in galactic nuclei relativistic precession substantially
reduces---and can even completely quench---resonant relaxation close to the
loss cone. A more sophisticated treatment than the one leading to equation
(\ref{eq:lossres}) would be required to determine the disruption rate and
density profile when the loss cone is empty. Such an analysis is beyond the
goals of this investigation.

\subsection{Relaxation Time Estimates}

The non-resonant relaxation time in a stellar system may be written
(e.g., Binney \& Tremaine 1987, eq.~8-71)
\be
t^{\rm nr}_{\rm rel}=0.3{\sigma^3\over G^2m\rho\ln\Lambda},
\ee
where $\sigma$ is the one-dimensional velocity dispersion and $\rho$ is the
stellar density. Specializing to the case of stars in the potential of a black
hole of mass $M$, we may write $\sigma\simeq (GM/3r)^{1/2}$, $\rho\simeq
M_\star/({4\over 3}\pi r^3)$, and $\Lambda\simeq N$; thus
\begin{eqnarray}
\trel^{\rm nr}& \simeq & 0.3{M^{3/2}r^{3/2}\over G^{1/2}m^2N\ln
N}\nonumber \\ & =  & 
{4\times10^{10}\hbox{ yr}\over\ln N}\left(M\over M_\star\right)\left(M\over
10^8M_\odot\right)^{1/2}\left(1 M_\odot\over m\right)\left(r\over
1\pc\right)^{3/2}.
\label{eq:relnrr} 
\end{eqnarray} 
The resonant relaxation time (cf. \S\S~\ref{sec_nrrel}--\ref{sec_resrel}) is
\be
\trel^{\rm res}\simeq {\eta_{\rm s}^2\over\beta_{\rm s}^2}{M_\star\over M}
\,\trel^{\rm nr}\simeq 7\,{M_\star\over M}\,\trel^{\rm nr},
\label{eq:relrat}
\ee
where $\beta_{\rm s}$ and $\eta_{\rm s}$ are taken from Table~\ref{aveparas}.

\begin{deluxetable}{cccccccc}
\tablecaption{Characteristics of Nearby Massive Black Hole Candidates
  \label{galtab}}

\tablehead{
\colhead{Galaxy} & \colhead{$D$} & \colhead{$M$} &
  \colhead{$\theta(\mu=0.1)$} & \colhead{$\mu(0\farcs1)$\tablenotemark{a}} &
  \colhead{$\trel^{\rm res}(0\farcs1)$\tablenotemark{a}} &
  \colhead{$\trel^{\rm nr}(0\farcs1)$\tablenotemark{a}} &
  \colhead{$\theta_{\rm res}$} \\
\colhead{} & \colhead{(Mpc)} & \colhead{($\Msun$)} & \colhead{$(\arcsec)$} &
  \colhead{} & \colhead{(yr)} & \colhead{(yr)} & \colhead{$(\arcsec)$}
}

\startdata
Milky Way& 0.0085& $2\times 10^6$& 6& 0.001& ${2\times10^{8\phn}}$&
  ${3\times10^{10\phn}}$& 12\nl
M32& 0.7& $2\times 10^6$& 0.08& 0.2& $4\times 10^{9\phn}$& $3\times 10^{9\phn}$&
  0.1\nl
M31& 0.7& $3\times 10^7$& 0.5& 0.002& $2\times 10^{10}$& $1\times 10^{12}$&
  0.07 \nl
NGC 3377& 9.9& $8\times10^7$& 0.05& 0.3& $1\times10^{12}$& $3\times10^{11}$&
  0.005\nl
NGC 3115& 8.4& $2\times 10^9$& 0.5& 0.02& $2\times10^{12}$&
  $2\times 10^{13}$& 0.002\nl
M87& 15.3& $3\times 10^9$& 1.3& 0.001& $8\times 10^{12}$& $1\times 10^{15}$&
  0.001\nl
\enddata

\tablenotetext{a}{For the Galactic Center only, these values are given at
$1\farcs0$, not $0\farcs1$. }

\tablecomments{Estimated distance $D$ and black-hole mass $M$ are from
Kormendy \& Richstone (1995).  The quantity $\mu(r)$ is the fraction of the
total mass inside radius $r$ that resides in stars; $\mu\ll1$ implies that the
potential is nearly Keplerian and when $\mu\lesssim 0.1$ resonant relaxation
dominates non-resonant relaxation (eq.~\ref{eq:relrat}). The resonant and
non-resonant relaxation times $\trel^{\rm res}$ and $\trel^{\rm nr}$ are given
by equations (\ref{eq:relnrr}) and (\ref{eq:relrat}); $\theta_{\rm res}$ is
the apparent angular distance from the black hole at which the resonant
relaxation time is $10^{10}$ yr. Stellar density estimates for external
galaxies are based on Hubble Space Telescope photometry (Faber et al.\ 1996)
and assumed
mass-to-light ratios; densities for the Milky Way are based on near-infrared
photometry (Kent 1992) and an assumed core radius of 0.15 pc (Eckart et al.\
1993).  }
\end{deluxetable}

Table~\ref{galtab}
applies these estimates to several nearby galaxies that are believed
to harbor massive black holes (Kormendy \& Richstone 1995). The table lists
the resonant and non-resonant relaxation times $\trel^{\rm res}$ and
$\trel^{\rm nr}$, at $0\farcs1$ from the center for external galaxies
(roughly Hubble Space Telescope resolution) and at $1\farcs0$ for the
Galaxy.

We see that resonant relaxation generally does not influence the structure of
galaxies ($\trel^{\rm res}\gtrsim 10^{10}$ yr) at available resolutions,
except in a few Local Group members such as M32 and the Galaxy. M31 contains a
region in which $\trel^{\rm res}<10^{10}\;{\rm yr}<\trel^{\rm nr}$ that is
almost (!)  resolvable by HST.  At smaller distances from the black hole the
resonant relaxation time is shorter; crude extrapolations indicate that
$\trel^{\rm res}\lesssim 10^{9}$ yr at radii less than $0\farcs05$ in M32, at
$0\farcs01$ in M31, and at $0\farcs002$ in NGC 3115. The first two of these
could be resolved with proposed space-based interferometers.

\section{Discussion}

Resonant relaxation is the dominant source of angular momentum relaxation for
stellar systems in near-Keplerian potentials, and thus plays an important role
in determining the structure of stellar cusps around black holes in galactic
nuclei or globular clusters. Resonant relaxation enhances the angular momentum
relaxation rate by roughly the ratio of the mass of the black hole to the mass
in stars but does not affect the energy relaxation rate (more precisely, the
combination of actions $J_1-J_2$ is conserved, where $J_1$ is the radial
action and $J_2$ is the angular momentum).  

Resonant relaxation is also present in the harmonic potentials that
characterize constant-density cores, and may enhance the rate of angular
momentum relaxation in the cores of globular clusters. In constant-density
cores, resonant relaxation preserves the combination of actions $J_1-2J_2$.
This form of resonance is not likely to be important for elliptical galaxies,
which do not generally have constant-density cores (e.g., Gebhardt et
al. 1996). 

One might speculate that generic potentials contain (high-order) resonances
that are strong enough to support resonant relaxation. In this case the
angular momentum relaxation time would be much shorter than the energy
relaxation time throughout most of a galaxy. We suspect that this
speculation is not correct, since our $N$-body simulations in the isochrone
potential yield very similar relaxation rates for the energy and angular
momentum. Nevertheless, the presence of resonant relaxation is a reminder
that our understanding of relaxation in stellar systems is crude, and has not
been numerically verified under conditions ($N\sim 10^{11}$) found in real
galaxies. 

Resonant friction leads to growth or decay of the eccentricity of
massive objects orbiting in near-Kepler potentials, depending on whether the
star orbits are predominantly radial or tangential. Resonant friction can
strongly influence the orbital evolution of a binary black hole (at least if
the mass ratio of the binary is sufficiently far from unity). In radially
biased star clusters the eccentricity of the binary will grow, at a rate
faster than the decay of the orbital energy, at least if the friction is
dominated by cluster stars rather than unbound stars. The binary eccentricity
will grow until the resonant friction is quenched by relativistic precession,
at which point gravitational radiation may erode the energy of the binary
faster than non-resonant dynamical friction. The details of this evolution are
relevant to the merger rate of black holes, the gravitational-wave background,
the prevalence of binary black holes in AGNs, and the viability of massive
black holes as dark matter candidates (see Quinlan 1996 for references).

Resonant friction can also erode the inclination of a massive object in a
rotating stellar system. This process may be relevant to a star cluster in
which there is a massive accretion disk (Ostriker 1983; Syer et al.\ 1991);
resonant friction could accelerate the evolution of massive stars into
low-inclination orbits embedded in the accretion disk.

The analytical treatment of resonant relaxation that we have offered in
\S~\ref{sec_analyt} is only approximate. Accurate expressions for the resonant
and non-resonant relaxation rates in a given star cluster could be derived by
expanding the potential from a stellar orbit in action-angle variables. So far
this procedure has only been carried out for the dynamical friction component
(Weinberg 1986).  The relative simplicity of the diffusion coefficients that
describe non-resonant relaxation (e.g., Binney \& Tremaine 1987) is illusory
in near-Keplerian and other near-resonant potentials---except to describe
energy relaxation---since resonant relaxation is stronger, and depends more
sensitively on the structure of the stellar system. For order-of-magnitude
estimates we have used the formulae in \S~\ref{sec_analyt}, with the
dimensionless coefficients given in Table~\ref{aveparas}.

The estimates of tidal disruption rates in \S~\ref{sec_newtloss} suffer from
the absence of a consistent treatment of relativistic precession, which
detunes the Kepler resonance near the loss cone in galactic nuclei. However,
resonant relaxation is unlikely to increase substantially the tidal disruption
rate, which is mostly determined by the location of the critical radius
$r_{\rm crit}$, set by the angular momentum changes in a single orbital
period. For similar reasons, resonant relaxation will not greatly affect the
disruption rate resulting when the wandering (``Brownian motion'') of the
black hole from the center of the nucleus (Quinlan 1995) is taken into
account.  Thus tidal disruption appears to remain incapable of powering
typical AGNs; however, since the relativistic detuning is largely ineffective
at hole masses $\lesssim 10^7\;\Msun$ (eq.~\ref{eq:grlimit}), resonant effects
may offer modest improvements in the feasibility of disruption-dominated mass
loss in Seyferts and other nuclei containing low-level AGN activity, for which
the energy requirements are less severe. Similarly, resonant relaxation may
modestly enhance the rate of flares from tidally disrupted stars in nearby
galaxies with central black holes (Rees 1988).

The effectiveness of relativistic precession in disabling resonant relaxation
illustrates that general relativity can have dramatic physical consequences
even where the motion is predominantly Newtonian; in particular, the shape of
the density cusp inside $r_{\rm crit}$ can be strongly dependent on
relativistic precession. Thus resonant relaxation might one day be used to
show that the massive dark objects observed in galactic nuclei are indeed
black holes (or at least behave as such on a scale of $\sim 10^2$
Schwarzschild radii)---a conclusion which today must be reached by indirect
(albeit compelling) arguments.

The discussion in \S 3.1 also illustrates that the Fokker-Planck equation used
to describe non-resonant relaxation (e.g. Binney and Tremaine 1987) is not
always adequate to describe resonant relaxation. The Fokker-Planck equation
assumes that the fluctuating forces at different times and locations are
uncorrelated, i.e., that the correlation function of \S 1.1 has the form
$C_{ij}({\mbf r}_1,{\mbf r}_2,\tau)=K_{ij}({\mbf r}_1)\delta({\mbf r}_2- {\mbf
r}_1)\delta(\tau)$. This is a reasonable approximation for non-resonant
relaxation, which is dominated by close encounters (cf. \S 1.1).  In contrast,
the resonant forces are correlated over large spatial scales and over times
$\sim\tprec$. The inadequacy of the Fokker-Planck approximation (or the master
equation, or the approximation that relaxation is a Markov process), is
particularly acute in the diffusion limit (eq. \ref{eq:lossres}), when the
size of the loss cone in angular-momentum space $\Lmin$ is much greater than
the change in angular momentum per orbit $\delLo$ but much less than
the change in angular momentum per precession time.

There is an appealing analogy between relaxation of stars in angular-momentum
space and models of stellar structure. Non-resonant relaxation is a random
walk in $L$-space, as for the motion of ions in the radiative zone of a
star. Resonant relaxation implies a large-scale drift in $L$ superimposed upon
a small-scale random walk, analogous to ionic motion in a convective stellar
envelope. The quenching of resonant relaxation by relativistic precession can
produce a random-walk dominated ``core'' in $L$-space surrounded by a
drift-dominated envelope---conceptually similar to the radiative
core/convective envelope structure found in solar-type stars. There are
fundamental differences: $L$-space has no analog to gravity, but there 
{\em is\/} a net flux of stars towards small $L$ due to removal of stars by
tidal disruption.

The structure of a relaxed star cluster surrounding a black hole has been
examined by several authors (Peebles 1972; Bahcall \& Wolf 1976, 1977;
Lightman \& Shapiro 1977; Cohn \& Kulsrud 1978). These analyses do not take
resonant relaxation into account and therefore some of their conclusions may
be suspect: we expect that including resonant relaxation will not strongly
affect the structure of the star cluster outside the critical radius $r_{\rm
crit}$ (\S~\ref{sec_newtloss}) or the total flux of stars into the loss cone,
but may substantially reduce the density of stars inside $r_{\rm crit}$. This
classic problem should be re-investigated. 

Resonant relaxation implies that there may be regions near the centers of
elliptical galaxies (typically $\lesssim 1\pc$ in radial extent; cf.
Table~\ref{galtab}) that are relaxed in angular momentum but not energy. If
non-rotating, such regions will have isotropic distribution functions; if
rotating, the mean rotation speed will depend on the stellar mass.
Unfortunately, these regions are not accessible at Hubble Space Telescope
resolutions in most nearby galaxies. A more fundamental problem regarding the
possible observational detection of resonant relaxation is that the isotropy
it produces will be undetectable unless the initial distribution function is
significantly anisotropic; there is currently no evidence for this in observed
nuclear star clusters.

\acknowledgments

We thank Jihad Touma for supplying the program to compute the average torque
between two Keplerian orbits. This research was supported by NSERC, and by a
Jeffrey L. Bishop Fellowship to K. R.

\end{document}